\documentclass[pra,twocolumn,showpacs,twoside]{revtex4-1}
\usepackage{graphicx}
\usepackage{times}
\usepackage{color}

\newcommand{\Erec}{E_\mathrm{rec}}
\newcommand{\rmax}{r_\mathrm{max}}
\newcommand{\pmax}{p_\mathrm{max}}
\newcommand{\kr}{k_\mathrm{r}}
\newcommand{\llat}{\lambda_\mathrm{Lat}}
\newcommand{\mRb}{m_\mathrm{Rb}}
\newcommand{\tof}{t_\mathrm{TOF}}
\newcommand{\xtof}{x_\mathrm{TOF}}

\begin{document}
\title{Dynamical control of matter wave splitting using time-dependent optical lattices}
\author{Sung Jong Park}
\altaffiliation{Korea Research Institute of Standards and Science, Yuseong, Daejeon
305-340, Korea}
\author{Henrik Kjaer Andersen}
\author{Sune Mai}
\author{Jan Arlt}
\author{Jacob F. Sherson}
\email{sherson@phys.au.dk}
\affiliation{Danish National Research Foundation Center for Quantum Optics, Department of Physics and Astronomy, University of Aarhus, DK-8000 Aarhus C, Denmark}

\date{\today}

\begin{abstract}
We report on measurements of splitting Bose-Einstein condensates (BEC) by using a time-dependent optical lattice potential. First, we demonstrate the division of a BEC into a set of equally populated components by means of time dependent control of  Landau-Zener tunneling in a vertical lattice potential. Next, we apply time dependent optical Bragg mirrors to a BEC oscillating in a harmonic trap. We demonstrate high-order Bragg reflection of the condensate due to multi-photon Raman transitions, where the depth of the optical lattice potential allows for a choice of the order of the transition. Finally, a combination of multiple Bragg reflections and Landau-Zener tunneling allows for the generation of macroscopic arrays of condensates with potential applications in atom optics and atom interferometry.

\end{abstract}
\pagestyle{plain}
\maketitle

\section{ Introduction}

The advent of Bose-Einstein condensates (BEC) as a source of coherent matter waves has revolutionized the field of atom-optics~\cite{Cronin2009}. In recent years, the understanding  has emerged that optical lattices provide a particularly rich tool for the manipulation of these coherent atomic ensembles~\cite{Morsch2006,Bloch2008}. Optical lattices allow e.g. for the realization of ordered and disordered lattice potentials, closely mimicking the situation in solid state materials~\cite{Lewenstein2007}. These systems can thus be viewed as quantum simulators for the solid state case. In addition, the ability to adjust the lattice parameters, even during a single experimental sequence, enables fundamentally new investigations in periodic potentials.

Experiments with quantum gases in optical lattices can roughly be divided into a static and a dynamic case. In the static case optical lattices have e.g. enabled fundamental investigations of many-body phase transitions~\cite{Greiner2002,Bakr:2010,Sherson:2010,Lewenstein2007} and allowed for interferometric combination of matter waves through tunneling~\cite{Anderson1998}.

In the dynamic case, pulses of optical lattice light have allowed for the manipulation of matter waves in the Kapitza-Dirac as well as Bragg regime. In the Kapitza-Dirac regime~\cite{Ovchinnikov1999}, a short pulse of optical lattice light projects between quasi-momentum and free momentum states of a Bose-Einstein condensate. The resulting atomic diffraction pattern is frequently used for lattice depth calibration~\cite{Denschlag2002,Morsch2006} and has recently been used to recombine matter waves interferometrically in free space~\cite{Hughes2009,Robert-de-Saint-Vincent2010} and in a harmonic confinement~\cite{Sapiro2009}.
When the interaction time between the lattice and the BEC is longer the Bragg regime is entered. In this regime the propagation within the band structure causes the appearance of numerous interesting physical phenomena. In particular when atoms are driven by a constant inertial force in an optical lattice, Bragg diffraction can occur~\cite{Kozuma1999} and atoms can undergo Bloch oscillations \cite{BenDahanM1996,Morsch2001}. When the edge of the Brillouin zone is reached during these oscillations both Bragg reflection and Landau-Zener tunneling \cite{Zener1932} can take place, depending on the size of the band gap. In the case of BECs accelerated from a stationary state, atoms can be Bragg reflected at the edge of the first Brillouin zone involving a two-photon Raman process, or a  Landau-Zener transition from the lowest band to the first excited band can occur. Higher order Bragg reflection involving a higher-order Raman process has also been observed using moving standing wave pulses~\cite{Kozuma1999,Denschlag2002} and Fourier-synthesized optical lattices~\cite{Ritt2006}. Moreover Landau-Zener tunneling has been used as a tool to measure the optical lattice depth as well as the effects of the mean-field interaction between the atoms in the condensate~\cite{Cristiani2002}. Recently, long lived Bloch oscillations of condensates in a vertical lattice have been achieved by suppressing the atomic interactions~\cite{Fattori2008, Gustavsson2008} wich provides a novel avenue for atom interferometry.

Although both the ground state and the dynamical properties of quantum gases have been investigated in detail, to date, only relatively few experiments fully exploit the possibility of dynamically controlling the lattice properties synchronously with the evolution of the quantum state.


Within the experiments presented here this is realized by dynamically controlling the depth of an optical lattice in synchrony with the evolution of a Bose-Einstein condensate. We investigate two distinct scenarios. In a first experiment we initiate Bloch oscillations in a vertical lattice under the constant force of gravity. Time-dependent control of the Landau-Zener tunneling rate enables us to realize a controlled matter wave beam splitter and a coherent matter wave source with controlled output coupler. In a second set of experiments we investigate the coherent splitting of a Bose-Einstein condensate  by pulsing on an optical lattice, corresponding to a time-dependent partial Bragg mirrors, in the presence of an external magnetic trapping potential

This allows us to investigate a matter wave splitter based on high-order Bragg reflection. It also enables the production of arrays of BECs by taking advantage of the anharmonicity of our trap. Although this impedes the interferometric recombination of BECs in our particular apparatus, the method may in the future lead to new interferometric avenues. This work thus extends the toolbox of splitting and recombination techniques from Kapitza-Dirac scattering~\cite{Hughes2009,Robert-de-Saint-Vincent2010,Sapiro2009} to the well-controlled matter wave manipulation with Bloch oscillations.

The paper is organized as follows. In Sec.~\ref{sec:setup} we briefly introduce our experimental setup. Section~\ref{sec:theo} reviews Bragg reflection and Landau-Zener tunneling in a periodic potential. The experimental results of time dependent Landau-Zener tunneling in free space are presented in Sec.~\ref{sec:exp1}. Finally Sec.~\ref{sec:exp2} introduces matter wave splitting and high-order Bragg reflection in the harmonic trap.

\section{experimental setup}
\label{sec:setup}
The experimental setup has been described in Ref.~\cite{Bertelsen2007} and only a brief summary is given here. Initially about $10^9$ $^{87}$Rb atoms are captured and cooled in a magneto-optical trap, subsequently these atoms are transfered into a magnetic quadrupole trap and transported mechanically to a science chamber. There, further evaporative cooling is performed in a quadrupole-Ioffe configuration trap (QUIC)~\cite{Esslinger-1998} and Bose-Einstein condensates with typically about $3 \times 10^5$ $^{87}$Rb atoms in the $F=2, m_F=2$ state are obtained. For optimal overlap with the optical lattice the magnetic trap is decompressed yielding trap frequencies of 12.3~Hz and 38.4~Hz in the axial and radial direction respectively.

The optical lattice is derived from a single mode diode laser operating at $\llat = 914$ nm, subsequently amplified in a tapered amplifier. The lattice is formed along the vertical direction by a retroreflected beam~($1/e^2$ waist of 120 $\mu$m). Atoms are loaded from the magnetic trap by adiabatically ramping up the lattice potential in a 100ms s-shaped ramp. The lattice depth can be varied from $0-30\Erec$, where $\Erec=h^2/(2m\llat^2)$, and adjusted with a time resolution on the order of $\mu $s.

Following the experiments, the condensate is released and resonant absorption images are taken after a variable time-of-flight expansion $\tof$.

\section{Theoretical background}
\label{sec:theo}

The field of cold atomic gases in optical lattices has attracted immense theoretical attention over the past decade~\cite{Lewenstein2007,Jaksch2005}. Here, we limit our discussion to the relevant case of a Bose-Einstein condensate in a one dimensional optical lattice with particular emphasis on the phenomena of Bloch oscillations, Bragg reflection and Landau-Zener tunneling.

The dynamics of a condensate in an optical lattice is governed by the band structure of the periodic potential. As the lattice is raised, the quadratic energy spectrum of a free particle splits into bands which are labeled by the eigenenergies $E_n(q)$ with the eigenstates $|n,q\rangle$, where $n=0,1,2,3...$ denotes the band index and $q$ the atomic quasi-momentum. The bands are separated by energy gaps whose size depends on the lattice depth. As a consequence, the dynamics of an atom moving in such a potential is dramatically altered compared to the free case.

 When an atom is subject to a force along the lattice axis, the band structure can cause the atom to start oscillating instead of being constantly accelerated. These so-called Bloch oscillations occur, due to Bragg reflection at the edge of the Brillouin zone at quasi-momentum $\hbar \kr$, where  $\kr=2\pi/\llat$ is the wave number of the lattice. The Bloch period is given by $\tau_B=2\hbar \kr/\mRb a$, where $\mRb$ is the mass of a rubidium atom and $a$ is the acceleration. It corresponds to the time it takes for the atoms to be accelerated from one end of the Brillouin zone to the other. This oscillation can also be interpreted as a harmonic oscillation in a matter wave cavity. If the lattice beams are arranged to create a periodic potential along the vertical direction, a condensate can be held against gravity for several seconds in a sufficiently deep lattice potenti9al, while the atoms perform Bloch oscillations~\cite{Gustavsson2008}.

In addition to the oscillatory behavior, tunneling between bands can occur at the edge of the Brillouin zone. In particular when the lattice potential is reduced and the band gaps narrow, so-called Landau-Zener tunneling between Bloch bands starts to occur.
If the atom moves with an acceleration $a$ through the avoided crossing of the $E_{n-1}(q)$ and $E_{n}(q)$ bands, the tunneling probability is approximately given  by~\cite{Zener1932}
\begin{equation}
\label{LZT-rate:}
P_t(n) = \exp{\left(-\frac{a_c(n)}{a}\right)} =  \exp\left(-\frac{\pi\Delta_n^2}{4 n \kr a \hbar^2} \right)
\end{equation}
where we give the generalized expression for tunneling between bands $n$ and $n-1$ for later reference. Here $a_c(n)$ is the critical acceleration and $\Delta_n$ is the size of the band gap at the quasi-momentum $n \hbar \kr$. In the experiments described in Sec.~\ref{sec:exp1} the acceleration is given by gravity ($a=g$) and in Sec.~\ref{sec:exp2} the acceleration is determined by the slope of the harmonic potential at the  position of the atoms.

Atomic interactions can
modify this behavior, however, the work presented here
is restricted to the linear regime where the tunneling probability follows the Landau-Zener formula. Experimentally the tunneling probability is set by controlling the optical lattice depth and can be suppressed completely when the lattice depth is increased.

\begin{figure}[tb]
\includegraphics[width=8cm]{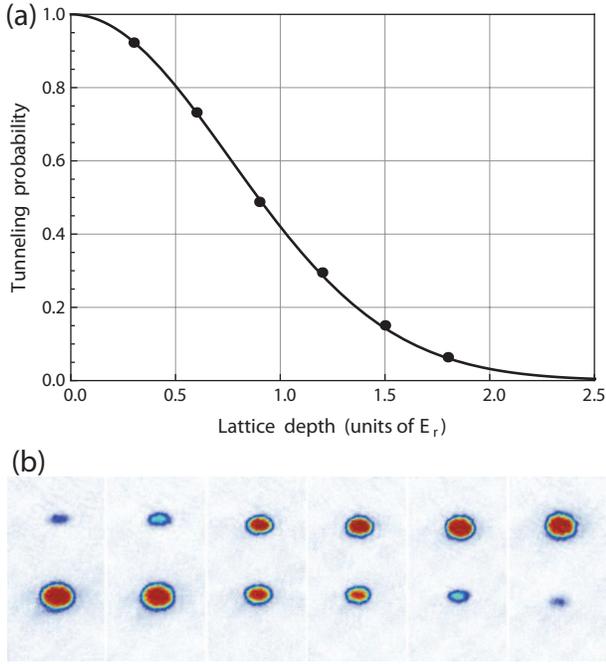}
\caption{(Color online). (a) Landau-Zener tunneling probability of a condensate between the two lowest energy bands in an optical lattice as a function of the lattice depth. The  solid line represents Eq.~(\ref{LZT-rate:}) and the  dots show the experimental results. The height of the dots indicates the representative standard deviation (3\%). (b) Experimental images corresponding to the upper graph.}
\label{LZTprobability}
\end{figure}

\section{Dynamical control of a matter wave beam splitter}
\label{sec:exp1}

In a first set of experiments we demonstrate that time-dependent control of the Landau-Zener tunneling rates enables the realization of  a controlled matter wave beam splitter and output coupler.

The splitting mechanism relies on the evolution of the atoms in the combined lattice and gravitational potential. The atoms start to perform Bloch oscillations and each time they reach the edge of the Brillouin zone a fraction of atoms can tunnel to a higher band. If the lattice depth is chosen such that atoms in higher bands with $n\geq1$ are not bound, these atoms start to fall under the influence of gravity. Thus the condensate is split each time the edge of the Brillouin zone is encountered and atoms with $n\geq1$ fall out of the lattice while Bragg reflected atoms remain trapped. Figure~\ref{LZTprobability} shows the measured Landau-Zener tunneling probability as a function of lattice depth, obtained from the fraction of tunneled to remaining atoms after a single Bloch oscillation. The very good agreement with Landau-Zener theory based on an independent measurement of the lattice depth confirms that interaction effects can be neglected.

If the tunneling probability is held fixed and the initial atom number is $N_0$ the number of atom tunneling out of the lattice on the $n$'th Bloch period is $N_0 (1-P_t(1))^{n-1}P_t(1)$ and $N_0(1-P_t(1))^n$ atoms remain in the lattice. This mechanism was indeed used in the first experiments with Bose-Einstein condensates in optical lattices~\cite{Anderson1998} and later investigated in detail~\cite{Morsch2001,Cristiani2002,Jona-Lasinio2003}.

Alternatively however, the dynamical control available in experiments with optical lattices can be used to taylor the outcoupling. If the lattice depth is controlled synchronously with the Bloch oscillation period, the tunneling rate can be controlled individually for each tunneling event and thus the fraction of outcoupled atoms can be determined at will.

\begin{figure}[tb]
\includegraphics[height=7.5cm]{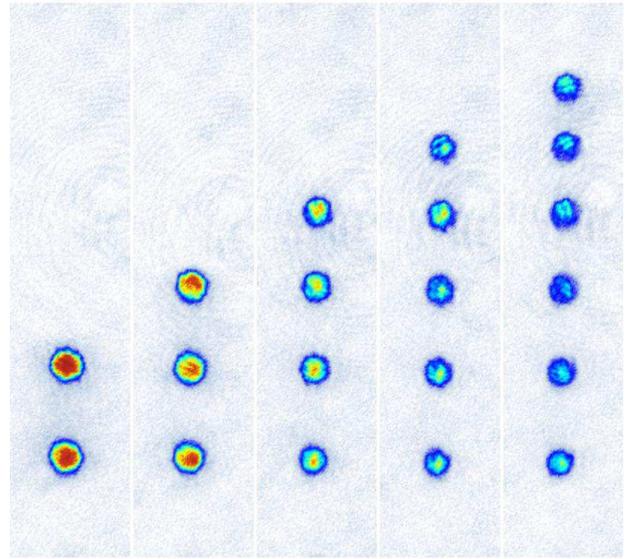}
\caption{(Color online). The condensate is divided into $n$ sub-condensates using time-dependent control of the tunneling probability. In order to produce $n$ sub-condensates with the same fraction, the tunneling probability is controlled by adjusting the lattice depth for each Bloch cycle. For example, tunneling probabilities of 1/6, 1/5, 1/4, 1/3, 1/2 are used to produce 6 clouds with the same size (standard deviation 6\%) as shown on the right.}
\label{fig:splitting-N}
\end{figure}

Within our experiments, the following sequence is used to employ this mechanism as a controllable beam splitter. After production of a Bose-Einstein condensate the magnetic trap is switched off to release the atoms. During  this process the atoms receive a small initial upwards velocity $v=9~mm/s$ and reach $q=0$  after 920~$\mu$s due to gravity.
The Bloch oscillations are initiated by turning on the lattice at a depth sufficient to completely suppress Landau-Zener tunneling (typically 2.4~$E_r$) just before the atoms reach the Bragg momentum ($\hbar \kr$). The lattice depth is ramped up within $200 \mu$s which is fast compared to the Bloch period of $\tau_B=1.0$~ms and sufficiently slow to ensure that the free space momentum is mapped to the corresponding quasi-momentum. To tailor the emission of matter wave packets we then adjust the lattice depth for each subsequent Bloch oscillation to enable controlled Landau-Zener tunneling based on Eq.~\ref{LZT-rate:}.

\begin{figure}[tb]
\includegraphics[height=7.5cm]{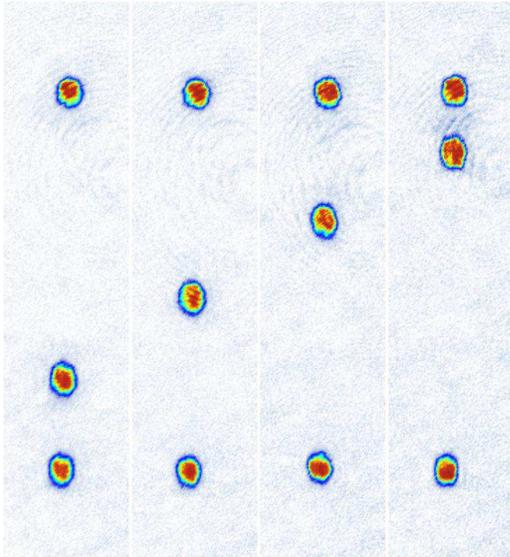}
\caption{(Color online). The condensate is divided into 3 sub-condensates which are produced within six Bloch cycles. The first (bottom) and last (top) wave-packets are released in Bloch oscillation cycle 1 and 6. The third one between them is released during cycle 2, 3, 4 or 5, while  tunneling  is completely suppressed for the remaining Bloch cycles.}
\label{fig:splitting-Z}
\end{figure}

Figure~\ref{fig:splitting-N} shows that the condensate can thus be divided into $m$ equally sized sub-condensates by varying the tunneling probability according to $1/m$, $1/(m-1)$, ..., $1/2$ for subsequent Bloch cycles and releasing the last one without splitting. The time-dependent control of the optical lattice depth can also be used to completely suppress the tunneling by increasing the lattice depth. Note that these images allow for a measurement of the tunneling probability and thus a calibration of the lattice depth with a single realization, rather than measuring the atom number that remains trapped after multiple Bloch periods $\tau_B$~\cite{Cristiani2002}.
In Fig.~\ref{fig:splitting-Z} we illustrate such controlled splitting of matter waves within six Bloch oscillation periods by releasing three matter wave packets with the same size while blocking the tunneling for the remaining three Bloch periods.

The precise control of the emission of matter wave packets demonstrated here is available for several tens of Bloch oscillations and is well suited as a beam splitter or outcoupler. In particular, as described below, it allows for the realization of large momentum transfer beam splitters~\cite{Denschlag2002,Clade2009} and may thus be useful for atom interferometry.
Moreover, the outcoupler presented here can be employed if a set of quasi-momenta are present in an optical lattice. In that case, produced e.g. by phase-matched scattering~\cite{Campbell2006, Hilligsoe2005, Gemelke2005}, a set of matter waves may oscillate in the lowest band with different phases, and each one can be emitted selectively with the time-dependent outcoupler. Hence the dynamical control of out-coupling can act as a phase-resolved matter wave beam splitter and thus provide a phase-matched coherent matter wave source.


\section{Matter wave splitting in harmonic confinement}
\label{sec:exp2}

In a second experiment  we investigate the coherent splitting of a BEC with an optical lattice in an external magnetic trapping potential. This method relies on precisely timed optical lattice pulses to split BECs using Bragg reflection and Landau-Zener tunneling during the oscillation in a harmonic trap.
In particular, this allows us to investigate a large momentum transfer matter wave splitter based on high-order Bragg reflection and enables the  production of arrays of trapped BECs by multiple splitting using time-dependent partial Bragg mirrors. A related procedure has recently been realized using Kapitza-Dirac scattering, allowing the implementation of atom interferometry by subsequently recombining the split clouds~\cite{Sapiro2009}.

In sections \ref{sec:theo} and \ref{sec:exp1} we have discussed the dynamics of a BEC oscillating in the zeroth band as it is being accelerated across the band edge by the force of gravity. If, on the other hand, the condensate is decelerated from a high momentum, Landau-Zener tunneling between higher excited bands and higher order Bragg reflection can occur depending on the lattice depth. In an
external  harmonic trap, this process can occur when the condensate is decelerated while approaching one of the classical turning points.

\begin{figure}[tb]
\includegraphics[width=1\columnwidth]{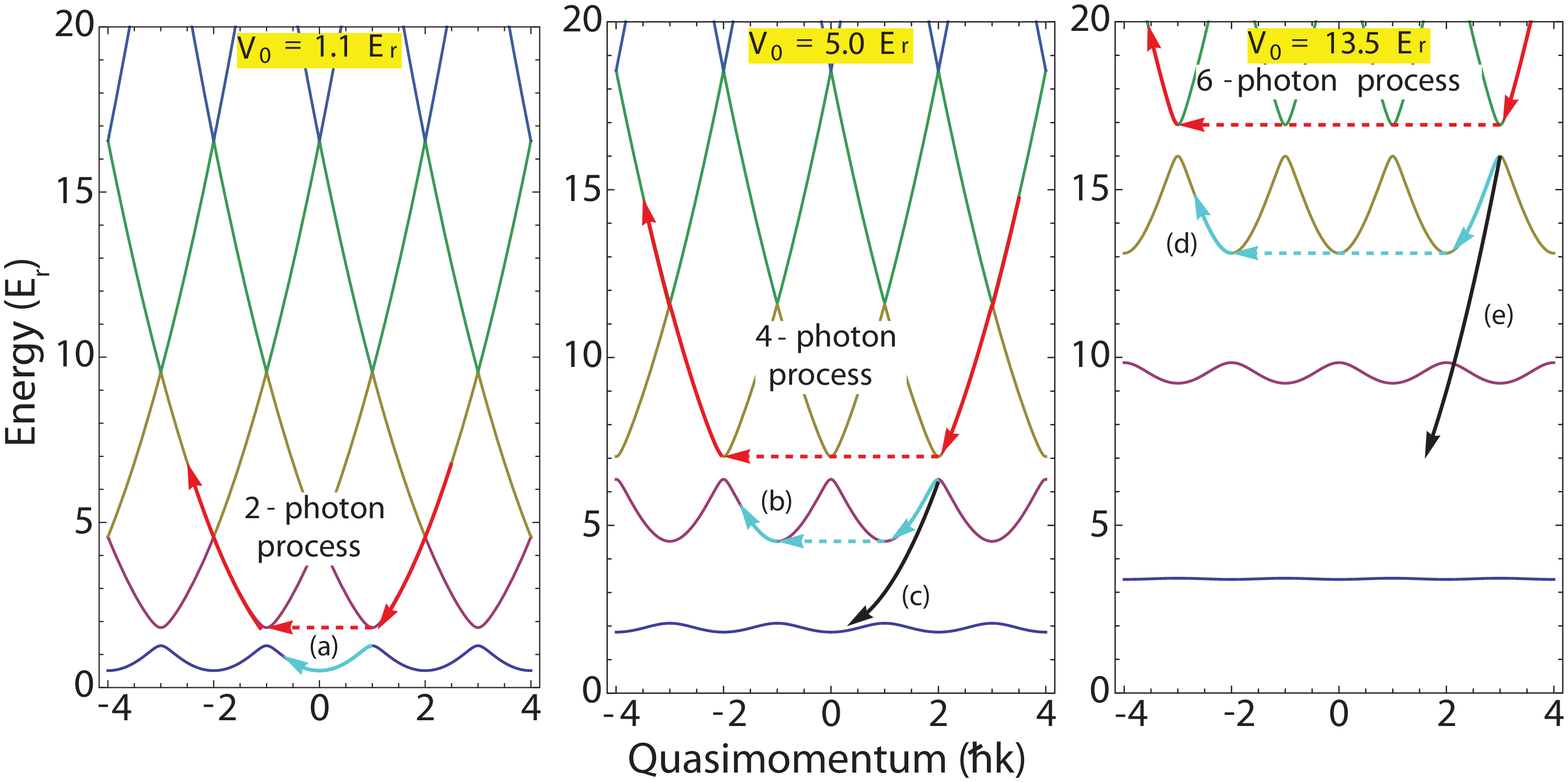}
\caption{(Color online). High-order Bragg reflection scheme. Half of the condensate is reflected (red arrows) on BM$_1$ (left), BM$_2$ (middle), and BM$_3$ (right), which correspond to  first-, second-, and third-order Bragg reflection, respectively. The transmitted fraction undergoes complete Bragg reflection at the following Brillouin zone boundary (light blue arrows) or propagates freely (black arrows) if the lattice is turned off. (a)-(e) refer to the individual pictures in Fig.~\ref{fig:refl-vs-depth-exp}.}
\label{fig:HO-reflection-scheme}
\end{figure}

Figure~\ref{fig:HO-reflection-scheme} illustrates this process in an extended Brillouin zone picture of the band structure for three different lattice depths. With increasing lattice depth, band gaps between higher bands appear, which in turn means that higher orders of Bragg reflection can occur. The lattice depths are chosen to have a tunneling probability of about 0.5 for the respective transition and thus realize a partially reflective Bragg mirror. We denote the partially reflecting Bragg mirror at the transition from band $n$ to band $n-1$ by BM$_n$.

For example, when a condensate is decelerated from large momentum at a lattice depth of 5~$E_r$, half of it is Bragg reflected from BM$_2$ while the other half is transmitted into the first band. This  fraction is then fully reflected at BM$_1$ as it reaches $q=\hbar\kr$ and hence stays in the first band. In real space this means that the Bragg reflected part reverses direction of propagation and is accelerated towards the trap center. The transmitted part continues to be decelerated and is  subsequently reflected on BM$_1$ before reaching the classical turning point.

In the following, two experimental investigations based on this mechanism are discussed.

\subsection{High order Bragg reflection}
\label{sec:HO-bragg-refl-exp}

Bragg reflection for atoms can also be described as a multi-photon Raman process, in which an atom with an initial momentum of $p=n \hbar\kr$, in the direction of one of the two light beams forming the lattice is transferred to a state with momentum $p=- n \hbar\kr$. The integer $n$ is the order of the Bragg reflection, and obviously $2n$ photons are exchanged between the atoms and the light field.

\begin{figure}[tb]
\centering\includegraphics[width=0.9\columnwidth]{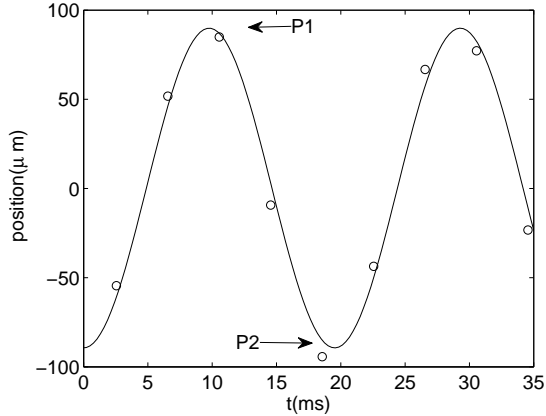}
\caption{Harmonic oscillation of a Bose-Einstein condensate. The dots show the measured \textit{in-situ} positions and the solid line is a fit with a frequency of 51.2 Hz and amplitude of 89.6 $\mu$m. P1 and P2 indicate the first and second turning points, respectively. The time-dependent optical lattices are pulsed on when the condensate is close to the turning points (see text).}
\label{fig:trap-frequency}
\end{figure}

To implement high momentum transfer using a harmonically oscillating BEC, the magnetic trap is suddenly shifted along the vertical direction by increasing the current to the QUIC coil. Due to  this displacement, the BEC undergoes harmonic motion with an amplitude of $\rmax=89.6$ $\mu$m and oscillation period $T=19.5$ ms as illustrated in Fig.~\ref{fig:trap-frequency}.  This displacement results in a maximum acquired momentum $\pmax= \mRb\omega\rmax \approx 6\hbar \kr$. For future reference we define the first and second turning points at $t=T/2$ and $t=T$ as P1 and P2, respectively. After initiating the harmonic motion of the condensate in the magnetic trap, we apply the lattice at varying times shortly before reaching P1. This allows us to implement an initial quasi-momentum in bands n=1,2, and 3.

Since the duration of the lattice  pulse is short compared to the total oscillation period, $T$, the dynamics  of each individual Bragg reflection occurs at a single  spatial point. This means that the force (which is spatially dependent in a harmonic trap) is constant for the duration of the lattice dynamics. Knowing this force enables us to adjust the lattice depth to implement a tunneling probability of about $P_t(n)=0.5$.

\begin{figure}[tb]
\includegraphics[width=0.9\columnwidth]{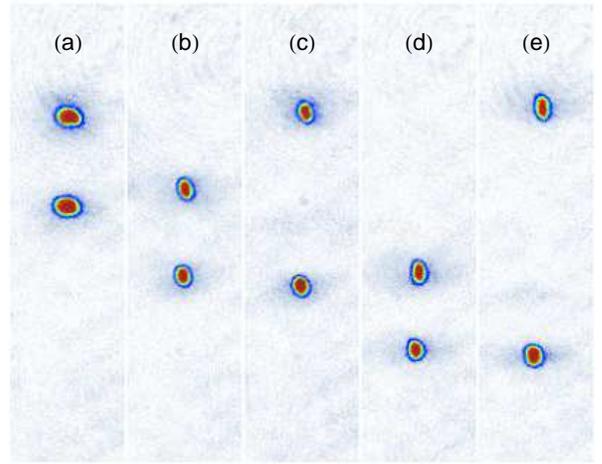}
\caption{(Color online). High-order Bragg reflection images. The condensate is split by the interaction with optical lattices with the depth of (a) 1.1 $E_r$, (b) and (c) 5.0 $E_r$, (d) and (e) 13.5 $E_r$.
See Fig.~\ref{fig:HO-reflection-scheme} for a graphical explanation of the distribution.}
\label{fig:refl-vs-depth-exp}
\end{figure}

Experimentally, we apply a $1$~ms lattice pulse as the momentum reaches 1, 2, and 3 $\hbar k$ for lattice depths of $1.1\Erec$, $5.0\Erec$, and $13.5\Erec$ respectively. For the lowest lattice depth (Fig. \ref{fig:refl-vs-depth-exp} a) this lattice pulse indeed results in a 50/50 splitting, followed by a brief period of propagation in the lattice for both clouds.
The release and subsequent time-of-flight ($\tof=12$ms) hence gives two clouds separated by $\xtof=\frac{2\hbar\kr}{\mRb}\tof$ \footnote{This is only approximately true, since the continued propagation in the harmonic confinement imparts different momentum corrections to each cloud.}.

For the deeper lattices, the first pulse also realizes a 50/50 splitting on $BM_n$ followed by a complete Bragg reflection of order $n-1$ for the part that tunneled to a lower band (light blue arrows in Fig.~\ref{fig:HO-reflection-scheme}). By turning off the lattice before the tunneled part reaches $q=-n\hbar\kr$ we again observe two peaks separated by $\xtof$ but shifted by approximately $n\xtof$ compared to the $1.1 E_r$ case. This can be seen in Fig. \ref{fig:refl-vs-depth-exp} b) and d), where the motion in the harmonic confinement in the time between $BM_n$ and $BM_{n-1}$ gives an offset (notice that $BM_n$ and $BM_{n-1}$ occur at very different points in the harmonic motion in the external potential).

The second, full Bragg reflection can be avoided by turning off the lattice before $q=(n-1)\hbar\kr$ is reached. In this case, illustrated in Fig. \ref{fig:refl-vs-depth-exp} c) and e), we observe a cloud unaffected by the lattice and one shifted by $n\xtof$ (black arrows in Fig.~\ref{fig:HO-reflection-scheme}).

This series of matter wave splitting demonstrates a large momentum transfer beam splitter (LMTBS), which was previously realized using accelerated optical lattices in Refs.~\cite{Denschlag2002,Clade2009}. Our experiments thus show that these higher-order Bragg reflections can be manipulated precisely by controlling the lattice depth and pulse timing.

\subsection{Arrays of trapped BECs}

An interesting application of the beam splitting process is the production of many independent BECs in the harmonic trap.
We realize this by pulsing on  Bragg mirrors before both turning points (P1 and P2) in the harmonic trap are reached.
To experimentally generate a prototype of macroscopic matter wave packet arrays, we first divide a BEC into one pair of matter wave packets which is separated by $2\hbar\kr$ momentum difference along the vertical axis, and then add a second partial Bragg mirror.

\begin{figure}[tb]
\centering\includegraphics[width=\columnwidth]{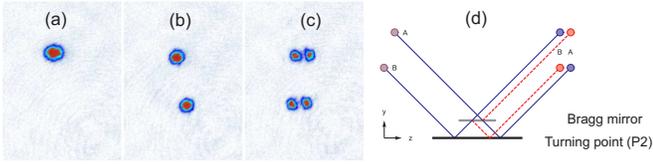}
\caption{(Color online). Production of wave packet arrays. (a) The initial BEC released after harmonic oscillation. (b) The BEC is split into two clouds after employing the Bragg mirror (BM$_1$) prior to the turning point P1. The two clouds are not vertically aligned due to the transverse motion (see text). (c) The split clouds are split again using the Bragg mirror (BM$_1$) at P2. Thus, pairs of sub-BECs are produced. (d) Schematic representation of the recombination process. }
\label{fig:Bragg-recomb-exp}
\end{figure}

Figure \ref{fig:Bragg-recomb-exp} shows the resulting matter wave packets after the individual steps of this procedure.
A Bragg mirror (BM$_1$) with tunneling probability of $P_t(1)=0.5$ is pulsed on briefly before the condensate reaches the turning point P1 to produce two equally split condensates with $2\hbar\kr$ momentum difference.
Figure \ref{fig:Bragg-recomb-exp} (b) shows the split wave-packets released following a reflection at the classical turning point   P2 (which in the terminology of this paper can be thought of as a perfect magnetic mirror).
Due to the gravitational shift, the principal axes tilt around the trap minimum~\cite{Modugno2003} and for the large oscillation amplitudes realized here, the mismatch between the local orientation of the principal axes of the potential and the vertical oscillation results in coupling into the transverse degrees of freedom.  This induced  motion is not harmonic and we have measured $\Delta z=3.5~\mu$m at $t=T/2$.
Figure \ref{fig:Bragg-recomb-exp}(b) shows this deviation.

In order to split the clouds again, we apply a second Bragg pulse in the vicinity of the turning point P2 at the time when the two clouds are spatially overlapped but have opposite momenta ($q\approx\pm\hbar\kr$). The Bragg reflected part of the first cloud now has the same momentum as the transmitted part of the second and vice versa.
Figure  \ref{fig:Bragg-recomb-exp}(c) shows the time-of-flight image obtained by turning off the magnetic field 2 ms after passing the second turning point P2.
As can be seen, the deviation from harmonicity of our trap allows for the production of arrays of distinguishable wave packets. The disadvantage is, however, that it is not possible to recombine the wave packets interferometrically.

\begin{figure}[tb]
\centering\includegraphics[width=\columnwidth]{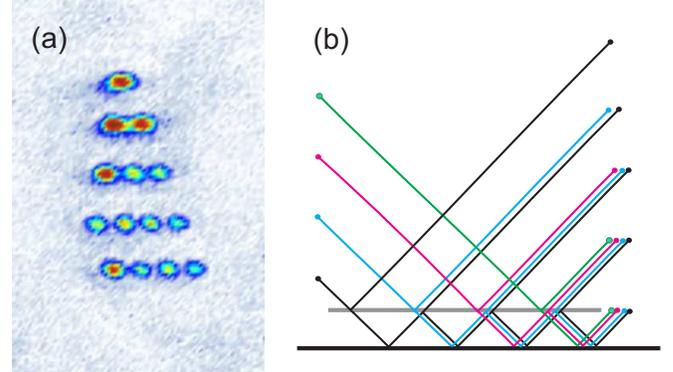}
\caption{(Color online). Production of arrays of matter waves. (a) Experimental image and (b) schematic representation of the recombination of four identical matter waves on partial Bragg- and perfect magnetic mirrors.}
\label{fig:Bragg-recomb-scheme}
\end{figure}

In a final experiment we extend this method to the production of large arrays of wave packets. To do this, we increase the duration of both lattice pulses above the Bloch oscillation time. First, a long lattice pulse produces an array of four condensates using BM$_1$  around P1 (see Sec. IV). Then, these clouds are split again by BM$_1$ around P2. As shown in Fig.~\ref{fig:Bragg-recomb-scheme}(a) we observe a pyramid-shaped arrays of condensates. The process is schematically represented in Fig.~\ref{fig:Bragg-recomb-scheme}(b).
Unlike the experiment described above, the second lattice pulse is on while all sub-clouds are moving towards the classical turning point. The part of the first cloud that was Bragg reflected is seen as the single cloud in the bottom row. The transmitted part continues and hits BM$_1$ simultaneously with the second cloud coming from P1 but with opposite momentum. The transmission of one and the reflection of the other explain the two clouds in the second row. Similarly, the third and fourth rows can be explained.

In principle an arbitrary amount of rows containing four atomic clouds with decreasing atom number can be produced by leaving the lattice on. In this demonstration, we turn off the lattice before the quadruple reaches the turning point again resulting in the images shown in Fig.~\ref{fig:Bragg-recomb-scheme}.

In this experiment the Bragg mirror acts as a partial mirror and the turning point  a perfect (magnetic) mirror. As can be seen in Fig \ref{fig:refl-vs-depth-exp}(b) and Fig. \ref{fig:refl-vs-depth-exp} (d), the magnetic mirror can be replaced by Bragg mirror. We have verified  that the same pyramidal arrays can be observed by increasing the lattice depth and never reaching the turning point.

\section{summary}

In summary, we have demonstrated coherent matter wave packet splitting using time-dependent control of the intensity of standing light waves. In this way the tunneling probability in each Bloch oscillation period can be precisely adjusted to control the outcoupled fraction and suppress tunneling completely with possible application as a precise matter wave source.

We have furthermore demonstrated the dynamics of a BEC in a combined potential consisting of a magnetic trap and an optical lattice. By setting the BEC in motion with a sudden displacement of the harmonic trap and pulsing on an optical lattice we have demonstrated complete control of high-order Bragg reflection of a BEC. Furthermore, we have produced matter wave packet arrays by pulsing on the  interaction with Bragg pulses  around  both turning points in a harmonic trap. This wave packet manipulation in an atom resonator resembles a misaligned Fabry-Perot resonator, where light enters and undergoes multiple internal reflections.

We acknowledge support from the Danish National Research Foundation and the Danish Council for Independent Research.

\bibliography{references}

\begin{thebibliography}{31}%
\makeatletter
\providecommand \@ifxundefined [1]{%
 \@ifx{#1\undefined}
}%
\providecommand \@ifnum [1]{%
 \ifnum #1\expandafter \@firstoftwo
 \else \expandafter \@secondoftwo
 \fi
}%
\providecommand \@ifx [1]{%
 \ifx #1\expandafter \@firstoftwo
 \else \expandafter \@secondoftwo
 \fi
}%
\providecommand \natexlab [1]{#1}%
\providecommand \enquote  [1]{``#1''}%
\providecommand \bibnamefont  [1]{#1}%
\providecommand \bibfnamefont [1]{#1}%
\providecommand \citenamefont [1]{#1}%
\providecommand \href@noop [0]{\@secondoftwo}%
\providecommand \href [0]{\begingroup \@sanitize@url \@href}%
\providecommand \@href[1]{\@@startlink{#1}\@@href}%
\providecommand \@@href[1]{\endgroup#1\@@endlink}%
\providecommand \@sanitize@url [0]{\catcode `\\12\catcode `\$12\catcode
  `\&12\catcode `\#12\catcode `\^12\catcode `\_12\catcode `\%12\relax}%
\providecommand \@@startlink[1]{}%
\providecommand \@@endlink[0]{}%
\providecommand \url  [0]{\begingroup\@sanitize@url \@url }%
\providecommand \@url [1]{\endgroup\@href {#1}{\urlprefix }}%
\providecommand \urlprefix  [0]{URL }%
\providecommand \Eprint [0]{\href }%
\providecommand \doibase [0]{http://dx.doi.org/}%
\providecommand \selectlanguage [0]{\@gobble}%
\providecommand \bibinfo  [0]{\@secondoftwo}%
\providecommand \bibfield  [0]{\@secondoftwo}%
\providecommand \translation [1]{[#1]}%
\providecommand \BibitemOpen [0]{}%
\providecommand \bibitemStop [0]{}%
\providecommand \bibitemNoStop [0]{.\EOS\space}%
\providecommand \EOS [0]{\spacefactor3000\relax}%
\providecommand \BibitemShut  [1]{\csname bibitem#1\endcsname}%
\let\auto@bib@innerbib\@empty
\bibitem [{\citenamefont {Cronin}\ \emph {et~al.}(2009)\citenamefont {Cronin},
  \citenamefont {Schmiedmayer},\ and\ \citenamefont {Pritchard}}]{Cronin2009}%
  \BibitemOpen
  \bibfield  {author} {\bibinfo {author} {\bibfnamefont {A.}~\bibnamefont
  {Cronin}}, \bibinfo {author} {\bibfnamefont {J.}~\bibnamefont
  {Schmiedmayer}}, \ and\ \bibinfo {author} {\bibfnamefont {D.}~\bibnamefont
  {Pritchard}},\ }\href {\doibase 10.1103/RevModPhys.81.1051} {\bibfield
  {journal} {\bibinfo  {journal} {Rev. Mod. Phys.}\ }\textbf {\bibinfo {volume}
  {81}},\ \bibinfo {pages} {1051} (\bibinfo {year} {2009})}\BibitemShut
  {NoStop}%
\bibitem [{\citenamefont {Morsch}\ and\ \citenamefont
  {Oberthaler}(2006)}]{Morsch2006}%
  \BibitemOpen
  \bibfield  {author} {\bibinfo {author} {\bibfnamefont {O.}~\bibnamefont
  {Morsch}}\ and\ \bibinfo {author} {\bibfnamefont {M.}~\bibnamefont
  {Oberthaler}},\ }\href {\doibase 10.1103/RevModPhys.78.179} {\bibfield
  {journal} {\bibinfo  {journal} {Rev. Mod. Phys.}\ }\textbf {\bibinfo {volume}
  {78}},\ \bibinfo {pages} {179} (\bibinfo {year} {2006})}\BibitemShut
  {NoStop}%
\bibitem [{\citenamefont {Bloch}\ and\ \citenamefont
  {Zwerger}(2008)}]{Bloch2008}%
  \BibitemOpen
  \bibfield  {author} {\bibinfo {author} {\bibfnamefont {I.}~\bibnamefont
  {Bloch}}\ and\ \bibinfo {author} {\bibfnamefont {W.}~\bibnamefont
  {Zwerger}},\ }\href {\doibase 10.1103/RevModPhys.80.885} {\bibfield
  {journal} {\bibinfo  {journal} {Rev. Mod. Phys.}\ }\textbf {\bibinfo {volume}
  {80}},\ \bibinfo {pages} {885} (\bibinfo {year} {2008})}\BibitemShut
  {NoStop}%
\bibitem [{\citenamefont {Lewenstein}\ \emph {et~al.}(2007)\citenamefont
  {Lewenstein}, \citenamefont {Sanpera}, \citenamefont {Ahufinger},
  \citenamefont {Damski}, \citenamefont {Sen(De)},\ and\ \citenamefont
  {Sen}}]{Lewenstein2007}%
  \BibitemOpen
  \bibfield  {author} {\bibinfo {author} {\bibfnamefont {M.}~\bibnamefont
  {Lewenstein}}, \bibinfo {author} {\bibfnamefont {A.}~\bibnamefont {Sanpera}},
  \bibinfo {author} {\bibfnamefont {V.}~\bibnamefont {Ahufinger}}, \bibinfo
  {author} {\bibfnamefont {B.}~\bibnamefont {Damski}}, \bibinfo {author}
  {\bibfnamefont {A.}~\bibnamefont {Sen(De)}}, \ and\ \bibinfo {author}
  {\bibfnamefont {U.}~\bibnamefont {Sen}},\ }\href {\doibase
  10.1080/00018730701223200} {\bibfield  {journal} {\bibinfo  {journal}
  {Advances in Physics}\ }\textbf {\bibinfo {volume} {56}},\ \bibinfo {pages}
  {243} (\bibinfo {year} {2007})}\BibitemShut {NoStop}%
\bibitem [{\citenamefont {Greiner}\ \emph {et~al.}(2002)\citenamefont
  {Greiner}, \citenamefont {Mandel}, \citenamefont {Esslinger}, \citenamefont
  {H\"{a}nsch},\ and\ \citenamefont {Bloch}}]{Greiner2002}%
  \BibitemOpen
  \bibfield  {author} {\bibinfo {author} {\bibfnamefont {M.}~\bibnamefont
  {Greiner}}, \bibinfo {author} {\bibfnamefont {O.}~\bibnamefont {Mandel}},
  \bibinfo {author} {\bibfnamefont {T.}~\bibnamefont {Esslinger}}, \bibinfo
  {author} {\bibfnamefont {T.~W.}\ \bibnamefont {H\"{a}nsch}}, \ and\ \bibinfo
  {author} {\bibfnamefont {I.}~\bibnamefont {Bloch}},\ }\href {\doibase
  10.1038/415039a} {\bibfield  {journal} {\bibinfo  {journal} {Nature}\
  }\textbf {\bibinfo {volume} {415}},\ \bibinfo {pages} {39} (\bibinfo {year}
  {2002})}\BibitemShut {NoStop}%
\bibitem [{\citenamefont {Bakr}\ \emph {et~al.}(2010)\citenamefont {Bakr},
  \citenamefont {Peng}, \citenamefont {Tai}, \citenamefont {Ma}, \citenamefont
  {Simon}, \citenamefont {Gillen}, \citenamefont {F\"{o}lling}, \citenamefont
  {Pollet},\ and\ \citenamefont {Greiner}}]{Bakr:2010}%
  \BibitemOpen
  \bibfield  {author} {\bibinfo {author} {\bibfnamefont {W.~S.}\ \bibnamefont
  {Bakr}}, \bibinfo {author} {\bibfnamefont {A.}~\bibnamefont {Peng}}, \bibinfo
  {author} {\bibfnamefont {M.~E.}\ \bibnamefont {Tai}}, \bibinfo {author}
  {\bibfnamefont {R.}~\bibnamefont {Ma}}, \bibinfo {author} {\bibfnamefont
  {J.}~\bibnamefont {Simon}}, \bibinfo {author} {\bibfnamefont {J.~I.}\
  \bibnamefont {Gillen}}, \bibinfo {author} {\bibfnamefont {S.}~\bibnamefont
  {F\"{o}lling}}, \bibinfo {author} {\bibfnamefont {L.}~\bibnamefont {Pollet}},
  \ and\ \bibinfo {author} {\bibfnamefont {M.}~\bibnamefont {Greiner}},\ }\href
  {http://arxiv.org/PS\_cache/arxiv/pdf/1006/1006.0754v1.pdf} {\bibfield
  {journal} {\bibinfo  {journal} {Science}\ }\textbf {\bibinfo {volume}
  {329}},\ \bibinfo {pages} {547} (\bibinfo {year} {2010})},\ \Eprint
  {http://arxiv.org/abs/arXiv:1006.0754v1} {arXiv:arXiv:1006.0754v1}
  \BibitemShut {NoStop}%
\bibitem [{\citenamefont {Sherson}\ \emph {et~al.}(2010)\citenamefont
  {Sherson}, \citenamefont {Weitenberg}, \citenamefont {Endres}, \citenamefont
  {Cheneau}, \citenamefont {Bloch},\ and\ \citenamefont {Kuhr}}]{Sherson:2010}%
  \BibitemOpen
  \bibfield  {author} {\bibinfo {author} {\bibfnamefont {J.~F.}\ \bibnamefont
  {Sherson}}, \bibinfo {author} {\bibfnamefont {C.}~\bibnamefont {Weitenberg}},
  \bibinfo {author} {\bibfnamefont {M.}~\bibnamefont {Endres}}, \bibinfo
  {author} {\bibfnamefont {M.}~\bibnamefont {Cheneau}}, \bibinfo {author}
  {\bibfnamefont {I.}~\bibnamefont {Bloch}}, \ and\ \bibinfo {author}
  {\bibfnamefont {S.}~\bibnamefont {Kuhr}},\ }\href {\doibase
  10.1038/nature09378} {\bibfield  {journal} {\bibinfo  {journal} {Nature}\
  }\textbf {\bibinfo {volume} {467}},\ \bibinfo {pages} {68} (\bibinfo {year}
  {2010})}\BibitemShut {NoStop}%
\bibitem [{\citenamefont {Anderson}\ and\ \citenamefont
  {Kasevich}(1998)}]{Anderson1998}%
  \BibitemOpen
  \bibfield  {author} {\bibinfo {author} {\bibfnamefont {B.~P.}\ \bibnamefont
  {Anderson}}\ and\ \bibinfo {author} {\bibfnamefont {M.~A.}\ \bibnamefont
  {Kasevich}},\ }\href {\doibase 10.1126/science.282.5394.1686} {\bibfield
  {journal} {\bibinfo  {journal} {Science}\ }\textbf {\bibinfo {volume}
  {282}},\ \bibinfo {pages} {1686} (\bibinfo {year} {1998})}\BibitemShut
  {NoStop}%
\bibitem [{\citenamefont {Ovchinnikov}\ \emph {et~al.}(1999)\citenamefont
  {Ovchinnikov}, \citenamefont {M\"{u}ller}, \citenamefont {Doery},
  \citenamefont {Vredenbregt}, \citenamefont {Helmerson}, \citenamefont
  {Rolston},\ and\ \citenamefont {Phillips}}]{Ovchinnikov1999}%
  \BibitemOpen
  \bibfield  {author} {\bibinfo {author} {\bibfnamefont {Y.~B.}\ \bibnamefont
  {Ovchinnikov}}, \bibinfo {author} {\bibfnamefont {J.~H.}\ \bibnamefont
  {M\"{u}ller}}, \bibinfo {author} {\bibfnamefont {M.~R.}\ \bibnamefont
  {Doery}}, \bibinfo {author} {\bibfnamefont {E.~J.~D.}\ \bibnamefont
  {Vredenbregt}}, \bibinfo {author} {\bibfnamefont {K.}~\bibnamefont
  {Helmerson}}, \bibinfo {author} {\bibfnamefont {S.~L.}\ \bibnamefont
  {Rolston}}, \ and\ \bibinfo {author} {\bibfnamefont {W.~D.}\ \bibnamefont
  {Phillips}},\ }\href {\doibase 10.1103/PhysRevLett.83.284} {\bibfield
  {journal} {\bibinfo  {journal} {Phys. Rev. Lett.}\ }\textbf {\bibinfo
  {volume} {83}},\ \bibinfo {pages} {284} (\bibinfo {year} {1999})}\BibitemShut
  {NoStop}%
\bibitem [{\citenamefont {Denschlag}\ \emph {et~al.}(2002)\citenamefont
  {Denschlag}, \citenamefont {Simsarian}, \citenamefont {H\"{a}ffner},
  \citenamefont {McKenzie}, \citenamefont {Browaeys}, \citenamefont {Cho},
  \citenamefont {Helmerson}, \citenamefont {Rolston},\ and\ \citenamefont
  {Phillips}}]{Denschlag2002}%
  \BibitemOpen
  \bibfield  {author} {\bibinfo {author} {\bibfnamefont {J.~H.}\ \bibnamefont
  {Denschlag}}, \bibinfo {author} {\bibfnamefont {J.~E.}\ \bibnamefont
  {Simsarian}}, \bibinfo {author} {\bibfnamefont {H.}~\bibnamefont
  {H\"{a}ffner}}, \bibinfo {author} {\bibfnamefont {C.}~\bibnamefont
  {McKenzie}}, \bibinfo {author} {\bibfnamefont {A.}~\bibnamefont {Browaeys}},
  \bibinfo {author} {\bibfnamefont {D.}~\bibnamefont {Cho}}, \bibinfo {author}
  {\bibfnamefont {K.}~\bibnamefont {Helmerson}}, \bibinfo {author}
  {\bibfnamefont {S.~L.}\ \bibnamefont {Rolston}}, \ and\ \bibinfo {author}
  {\bibfnamefont {W.~D.}\ \bibnamefont {Phillips}},\ }\href@noop {} {\bibfield
  {journal} {\bibinfo  {journal} {J. Phys. B: At. Mol. Opt. Phys.}\ }\textbf
  {\bibinfo {volume} {35}},\ \bibinfo {pages} {3095} (\bibinfo {year}
  {2002})}\BibitemShut {NoStop}%
\bibitem [{\citenamefont {Hughes}\ \emph {et~al.}(2009)\citenamefont {Hughes},
  \citenamefont {Burke},\ and\ \citenamefont {Sackett}}]{Hughes2009}%
  \BibitemOpen
  \bibfield  {author} {\bibinfo {author} {\bibfnamefont {K.}~\bibnamefont
  {Hughes}}, \bibinfo {author} {\bibfnamefont {J.}~\bibnamefont {Burke}}, \
  and\ \bibinfo {author} {\bibfnamefont {C.}~\bibnamefont {Sackett}},\ }\href
  {\doibase 10.1103/PhysRevLett.102.150403} {\bibfield  {journal} {\bibinfo
  {journal} {Phys. Rev. Lett.}\ }\textbf {\bibinfo {volume} {102}},\ \bibinfo
  {pages} {150403} (\bibinfo {year} {2009})}\BibitemShut {NoStop}%
\bibitem [{\citenamefont {Robert-de Saint-Vincent}\ \emph
  {et~al.}(2010)\citenamefont {Robert-de Saint-Vincent}, \citenamefont
  {Brantut}, \citenamefont {Bord\'{e}}, \citenamefont {Aspect}, \citenamefont
  {Bourdel},\ and\ \citenamefont {Bouyer}}]{Robert-de-Saint-Vincent2010}%
  \BibitemOpen
  \bibfield  {author} {\bibinfo {author} {\bibfnamefont {M.}~\bibnamefont
  {Robert-de Saint-Vincent}}, \bibinfo {author} {\bibfnamefont {J.-P.}\
  \bibnamefont {Brantut}}, \bibinfo {author} {\bibfnamefont {C.~J.}\
  \bibnamefont {Bord\'{e}}}, \bibinfo {author} {\bibfnamefont {A.}~\bibnamefont
  {Aspect}}, \bibinfo {author} {\bibfnamefont {T.}~\bibnamefont {Bourdel}}, \
  and\ \bibinfo {author} {\bibfnamefont {P.}~\bibnamefont {Bouyer}},\ }\href
  {\doibase 10.1209/0295-5075/89/10002} {\bibfield  {journal} {\bibinfo
  {journal} {EPL (Europhysics Letters)}\ }\textbf {\bibinfo {volume} {89}},\
  \bibinfo {pages} {10002} (\bibinfo {year} {2010})}\BibitemShut {NoStop}%
\bibitem [{\citenamefont {Sapiro}\ \emph {et~al.}(2009)\citenamefont {Sapiro},
  \citenamefont {Zhang},\ and\ \citenamefont {Raithel}}]{Sapiro2009}%
  \BibitemOpen
  \bibfield  {author} {\bibinfo {author} {\bibfnamefont {R.}~\bibnamefont
  {Sapiro}}, \bibinfo {author} {\bibfnamefont {R.}~\bibnamefont {Zhang}}, \
  and\ \bibinfo {author} {\bibfnamefont {G.}~\bibnamefont {Raithel}},\ }\href
  {\doibase 10.1103/PhysRevA.79.043630} {\bibfield  {journal} {\bibinfo
  {journal} {Phys. Rev. A}\ }\textbf {\bibinfo {volume} {79}},\ \bibinfo
  {pages} {043630} (\bibinfo {year} {2009})}\BibitemShut {NoStop}%
\bibitem [{\citenamefont {Kozuma}\ \emph {et~al.}(1999)\citenamefont {Kozuma},
  \citenamefont {Deng}, \citenamefont {Hagley}, \citenamefont {Wen},
  \citenamefont {Lutwak}, \citenamefont {Helmerson}, \citenamefont {Rolston},\
  and\ \citenamefont {Phillips}}]{Kozuma1999}%
  \BibitemOpen
  \bibfield  {author} {\bibinfo {author} {\bibfnamefont {M.}~\bibnamefont
  {Kozuma}}, \bibinfo {author} {\bibfnamefont {L.}~\bibnamefont {Deng}},
  \bibinfo {author} {\bibfnamefont {E.~W.}\ \bibnamefont {Hagley}}, \bibinfo
  {author} {\bibfnamefont {J.}~\bibnamefont {Wen}}, \bibinfo {author}
  {\bibfnamefont {R.}~\bibnamefont {Lutwak}}, \bibinfo {author} {\bibfnamefont
  {K.}~\bibnamefont {Helmerson}}, \bibinfo {author} {\bibfnamefont {S.~L.}\
  \bibnamefont {Rolston}}, \ and\ \bibinfo {author} {\bibfnamefont {W.~D.}\
  \bibnamefont {Phillips}},\ }\href@noop {} {\bibfield  {journal} {\bibinfo
  {journal} {Phys. Rev. Lett.}\ }\textbf {\bibinfo {volume} {82}},\ \bibinfo
  {pages} {871} (\bibinfo {year} {1999})}\BibitemShut {NoStop}%
\bibitem [{\citenamefont {Dahan}\ \emph {et~al.}(1996)\citenamefont {Dahan},
  \citenamefont {Peik}, \citenamefont {Reichel}, \citenamefont {Castin},\ and\
  \citenamefont {Salomon}}]{BenDahanM1996}%
  \BibitemOpen
  \bibfield  {author} {\bibinfo {author} {\bibfnamefont {B.~M.}\ \bibnamefont
  {Dahan}}, \bibinfo {author} {\bibfnamefont {E.}~\bibnamefont {Peik}},
  \bibinfo {author} {\bibfnamefont {J.}~\bibnamefont {Reichel}}, \bibinfo
  {author} {\bibfnamefont {Y.}~\bibnamefont {Castin}}, \ and\ \bibinfo {author}
  {\bibfnamefont {C.}~\bibnamefont {Salomon}},\ }\href
  {http://www.ncbi.nlm.nih.gov/pubmed/10061309} {\bibfield  {journal} {\bibinfo
   {journal} {Phys. Rev. Lett.}\ }\textbf {\bibinfo {volume} {76}},\ \bibinfo
  {pages} {4508} (\bibinfo {year} {1996})}\BibitemShut {NoStop}%
\bibitem [{\citenamefont {Morsch}\ \emph {et~al.}(2001)\citenamefont {Morsch},
  \citenamefont {M\"{u}ller}, \citenamefont {Cristiani}, \citenamefont
  {Ciampini},\ and\ \citenamefont {Arimondo}}]{Morsch2001}%
  \BibitemOpen
  \bibfield  {author} {\bibinfo {author} {\bibfnamefont {O.}~\bibnamefont
  {Morsch}}, \bibinfo {author} {\bibfnamefont {J.~H.}\ \bibnamefont
  {M\"{u}ller}}, \bibinfo {author} {\bibfnamefont {M.}~\bibnamefont
  {Cristiani}}, \bibinfo {author} {\bibfnamefont {D.}~\bibnamefont {Ciampini}},
  \ and\ \bibinfo {author} {\bibfnamefont {E.}~\bibnamefont {Arimondo}},\
  }\href {\doibase 10.1103/PhysRevLett.87.140402} {\bibfield  {journal}
  {\bibinfo  {journal} {Phys. Rev. Lett.}\ }\textbf {\bibinfo {volume} {87}},\
  \bibinfo {pages} {140402} (\bibinfo {year} {2001})}\BibitemShut {NoStop}%
\bibitem [{\citenamefont {Zener}(1932)}]{Zener1932}%
  \BibitemOpen
  \bibfield  {author} {\bibinfo {author} {\bibfnamefont {C.}~\bibnamefont
  {Zener}},\ }\href@noop {} {\bibfield  {journal} {\bibinfo  {journal} {Proc.
  R. Soc. London A}\ }\textbf {\bibinfo {volume} {173}},\ \bibinfo {pages}
  {696} (\bibinfo {year} {1932})}\BibitemShut {NoStop}%
\bibitem [{\citenamefont {Ritt}\ \emph {et~al.}(2006)\citenamefont {Ritt},
  \citenamefont {Geckeler}, \citenamefont {Salger}, \citenamefont {Cennini},\
  and\ \citenamefont {Weitz}}]{Ritt2006}%
  \BibitemOpen
  \bibfield  {author} {\bibinfo {author} {\bibfnamefont {G.}~\bibnamefont
  {Ritt}}, \bibinfo {author} {\bibfnamefont {C.}~\bibnamefont {Geckeler}},
  \bibinfo {author} {\bibfnamefont {T.}~\bibnamefont {Salger}}, \bibinfo
  {author} {\bibfnamefont {G.}~\bibnamefont {Cennini}}, \ and\ \bibinfo
  {author} {\bibfnamefont {M.}~\bibnamefont {Weitz}},\ }\href {\doibase
  10.1103/PhysRevA.74.063622} {\bibfield  {journal} {\bibinfo  {journal} {Phys.
  Rev. A}\ }\textbf {\bibinfo {volume} {74}},\ \bibinfo {pages} {063622}
  (\bibinfo {year} {2006})}\BibitemShut {NoStop}%
\bibitem [{\citenamefont {Cristiani}\ \emph {et~al.}(2002)\citenamefont
  {Cristiani}, \citenamefont {Morsch}, \citenamefont {M\"{u}ller},
  \citenamefont {Ciampini},\ and\ \citenamefont {Arimondo}}]{Cristiani2002}%
  \BibitemOpen
  \bibfield  {author} {\bibinfo {author} {\bibfnamefont {M.}~\bibnamefont
  {Cristiani}}, \bibinfo {author} {\bibfnamefont {O.}~\bibnamefont {Morsch}},
  \bibinfo {author} {\bibfnamefont {J.}~\bibnamefont {M\"{u}ller}}, \bibinfo
  {author} {\bibfnamefont {D.}~\bibnamefont {Ciampini}}, \ and\ \bibinfo
  {author} {\bibfnamefont {E.}~\bibnamefont {Arimondo}},\ }\href {\doibase
  10.1103/PhysRevA.65.063612} {\bibfield  {journal} {\bibinfo  {journal} {Phys.
  Rev. A}\ }\textbf {\bibinfo {volume} {65}},\ \bibinfo {pages} {063612}
  (\bibinfo {year} {2002})}\BibitemShut {NoStop}%
\bibitem [{\citenamefont {Fattori}\ \emph {et~al.}(2008)\citenamefont
  {Fattori}, \citenamefont {D'Errico}, \citenamefont {Roati}, \citenamefont
  {Zaccanti}, \citenamefont {Jona-Lasinio}, \citenamefont {Modugno},
  \citenamefont {Inguscio},\ and\ \citenamefont {Modugno}}]{Fattori2008}%
  \BibitemOpen
  \bibfield  {author} {\bibinfo {author} {\bibfnamefont {M.}~\bibnamefont
  {Fattori}}, \bibinfo {author} {\bibfnamefont {C.}~\bibnamefont {D'Errico}},
  \bibinfo {author} {\bibfnamefont {G.}~\bibnamefont {Roati}}, \bibinfo
  {author} {\bibfnamefont {M.}~\bibnamefont {Zaccanti}}, \bibinfo {author}
  {\bibfnamefont {M.}~\bibnamefont {Jona-Lasinio}}, \bibinfo {author}
  {\bibfnamefont {M.}~\bibnamefont {Modugno}}, \bibinfo {author} {\bibfnamefont
  {M.}~\bibnamefont {Inguscio}}, \ and\ \bibinfo {author} {\bibfnamefont
  {G.}~\bibnamefont {Modugno}},\ }\href {\doibase
  10.1103/PhysRevLett.100.080405} {\bibfield  {journal} {\bibinfo  {journal}
  {Phys. Rev. Lett.}\ }\textbf {\bibinfo {volume} {100}},\ \bibinfo {pages}
  {080405} (\bibinfo {year} {2008})}\BibitemShut {NoStop}%
\bibitem [{\citenamefont {Gustavsson}\ \emph {et~al.}(2008)\citenamefont
  {Gustavsson}, \citenamefont {Haller}, \citenamefont {Mark}, \citenamefont
  {Danzl}, \citenamefont {Rojas-Kopeinig},\ and\ \citenamefont
  {N\"{a}gerl}}]{Gustavsson2008}%
  \BibitemOpen
  \bibfield  {author} {\bibinfo {author} {\bibfnamefont {M.}~\bibnamefont
  {Gustavsson}}, \bibinfo {author} {\bibfnamefont {E.}~\bibnamefont {Haller}},
  \bibinfo {author} {\bibfnamefont {M.}~\bibnamefont {Mark}}, \bibinfo {author}
  {\bibfnamefont {J.}~\bibnamefont {Danzl}}, \bibinfo {author} {\bibfnamefont
  {G.}~\bibnamefont {Rojas-Kopeinig}}, \ and\ \bibinfo {author} {\bibfnamefont
  {H.-C.}\ \bibnamefont {N\"{a}gerl}},\ }\href {\doibase
  10.1103/PhysRevLett.100.080404} {\bibfield  {journal} {\bibinfo  {journal}
  {Phys. Rev. Lett.}\ }\textbf {\bibinfo {volume} {100}},\ \bibinfo {pages}
  {080404} (\bibinfo {year} {2008})}\BibitemShut {NoStop}%
\bibitem [{\citenamefont {Bertelsen}\ \emph {et~al.}(2007)\citenamefont
  {Bertelsen}, \citenamefont {Andersen}, \citenamefont {Mai},\ and\
  \citenamefont {Budde}}]{Bertelsen2007}%
  \BibitemOpen
  \bibfield  {author} {\bibinfo {author} {\bibfnamefont {J.}~\bibnamefont
  {Bertelsen}}, \bibinfo {author} {\bibfnamefont {H.}~\bibnamefont {Andersen}},
  \bibinfo {author} {\bibfnamefont {S.}~\bibnamefont {Mai}}, \ and\ \bibinfo
  {author} {\bibfnamefont {M.}~\bibnamefont {Budde}},\ }\href {\doibase
  10.1103/PhysRevA.75.013404} {\bibfield  {journal} {\bibinfo  {journal} {Phys.
  Rev. A}\ }\textbf {\bibinfo {volume} {75}},\ \bibinfo {pages} {013404}
  (\bibinfo {year} {2007})}\BibitemShut {NoStop}%
\bibitem [{\citenamefont {Esslinger}\ \emph {et~al.}(1998)\citenamefont
  {Esslinger}, \citenamefont {Bloch},\ and\ \citenamefont
  {H\"{a}nsch}}]{Esslinger-1998}%
  \BibitemOpen
  \bibfield  {author} {\bibinfo {author} {\bibfnamefont {T.}~\bibnamefont
  {Esslinger}}, \bibinfo {author} {\bibfnamefont {I.}~\bibnamefont {Bloch}}, \
  and\ \bibinfo {author} {\bibfnamefont {T.~W.}\ \bibnamefont {H\"{a}nsch}},\
  }\href@noop {} {\bibfield  {journal} {\bibinfo  {journal} {Phys. Rev. A}\
  }\textbf {\bibinfo {volume} {58}},\ \bibinfo {pages} {R2664} (\bibinfo {year}
  {1998})}\BibitemShut {NoStop}%
\bibitem [{\citenamefont {Jaksch}\ and\ \citenamefont
  {Zoller}(2005)}]{Jaksch2005}%
  \BibitemOpen
  \bibfield  {author} {\bibinfo {author} {\bibfnamefont {D.}~\bibnamefont
  {Jaksch}}\ and\ \bibinfo {author} {\bibfnamefont {P.}~\bibnamefont
  {Zoller}},\ }\href {\doibase 10.1016/j.aop.2004.09.010} {\bibfield  {journal}
  {\bibinfo  {journal} {Annals of Physics}\ }\textbf {\bibinfo {volume}
  {315}},\ \bibinfo {pages} {52} (\bibinfo {year} {2005})}\BibitemShut
  {NoStop}%
\bibitem [{\citenamefont {Jona-Lasinio}\ \emph {et~al.}(2003)\citenamefont
  {Jona-Lasinio}, \citenamefont {Morsch}, \citenamefont {Cristiani},
  \citenamefont {Malossi}, \citenamefont {M\"{u}ller}, \citenamefont
  {Courtade}, \citenamefont {Anderlini},\ and\ \citenamefont
  {Arimondo}}]{Jona-Lasinio2003}%
  \BibitemOpen
  \bibfield  {author} {\bibinfo {author} {\bibfnamefont {M.}~\bibnamefont
  {Jona-Lasinio}}, \bibinfo {author} {\bibfnamefont {O.}~\bibnamefont
  {Morsch}}, \bibinfo {author} {\bibfnamefont {M.}~\bibnamefont {Cristiani}},
  \bibinfo {author} {\bibfnamefont {N.}~\bibnamefont {Malossi}}, \bibinfo
  {author} {\bibfnamefont {J.}~\bibnamefont {M\"{u}ller}}, \bibinfo {author}
  {\bibfnamefont {E.}~\bibnamefont {Courtade}}, \bibinfo {author}
  {\bibfnamefont {M.}~\bibnamefont {Anderlini}}, \ and\ \bibinfo {author}
  {\bibfnamefont {E.}~\bibnamefont {Arimondo}},\ }\href {\doibase
  10.1103/PhysRevLett.91.230406} {\bibfield  {journal} {\bibinfo  {journal}
  {Phys. Rev. Lett.}\ }\textbf {\bibinfo {volume} {91}},\ \bibinfo {pages}
  {230406} (\bibinfo {year} {2003})}\BibitemShut {NoStop}%
\bibitem [{\citenamefont {Clad\'{e}}\ \emph {et~al.}(2009)\citenamefont
  {Clad\'{e}}, \citenamefont {Guellati-Kh\'{e}lifa}, \citenamefont {Nez},\ and\
  \citenamefont {Biraben}}]{Clade2009}%
  \BibitemOpen
  \bibfield  {author} {\bibinfo {author} {\bibfnamefont {P.}~\bibnamefont
  {Clad\'{e}}}, \bibinfo {author} {\bibfnamefont {S.}~\bibnamefont
  {Guellati-Kh\'{e}lifa}}, \bibinfo {author} {\bibfnamefont {F.}~\bibnamefont
  {Nez}}, \ and\ \bibinfo {author} {\bibfnamefont {F.}~\bibnamefont
  {Biraben}},\ }\href {\doibase 10.1103/PhysRevLett.102.240402} {\bibfield
  {journal} {\bibinfo  {journal} {Phys. Rev. Lett.}\ }\textbf {\bibinfo
  {volume} {102}},\ \bibinfo {pages} {240402} (\bibinfo {year}
  {2009})}\BibitemShut {NoStop}%
\bibitem [{\citenamefont {Campbell}\ \emph {et~al.}(2006)\citenamefont
  {Campbell}, \citenamefont {Mun}, \citenamefont {Boyd}, \citenamefont
  {Streed}, \citenamefont {Ketterle},\ and\ \citenamefont
  {Pritchard}}]{Campbell2006}%
  \BibitemOpen
  \bibfield  {author} {\bibinfo {author} {\bibfnamefont {G.~K.}\ \bibnamefont
  {Campbell}}, \bibinfo {author} {\bibfnamefont {J.}~\bibnamefont {Mun}},
  \bibinfo {author} {\bibfnamefont {M.}~\bibnamefont {Boyd}}, \bibinfo {author}
  {\bibfnamefont {E.~W.}\ \bibnamefont {Streed}}, \bibinfo {author}
  {\bibfnamefont {W.}~\bibnamefont {Ketterle}}, \ and\ \bibinfo {author}
  {\bibfnamefont {D.~E.}\ \bibnamefont {Pritchard}},\ }\href {\doibase
  10.1103/PhysRevLett.96.020406} {\bibfield  {journal} {\bibinfo  {journal}
  {Phys. Rev. Lett.}\ }\textbf {\bibinfo {volume} {96}},\ \bibinfo {pages}
  {020406} (\bibinfo {year} {2006})}\BibitemShut {NoStop}%
\bibitem [{\citenamefont {Hilligs\o~e}\ and\ \citenamefont
  {M\o~lmer}(2005)}]{Hilligsoe2005}%
  \BibitemOpen
  \bibfield  {author} {\bibinfo {author} {\bibfnamefont {K.~M.}\ \bibnamefont
  {Hilligs\o~e}}\ and\ \bibinfo {author} {\bibfnamefont {K.}~\bibnamefont
  {M\o~lmer}},\ }\href {\doibase 10.1103/PhysRevA.71.041602} {\bibfield
  {journal} {\bibinfo  {journal} {Phys. Rev. A}\ }\textbf {\bibinfo {volume}
  {71}},\ \bibinfo {pages} {041602(R)} (\bibinfo {year} {2005})}\BibitemShut
  {NoStop}%
\bibitem [{\citenamefont {Gemelke}\ \emph {et~al.}(2005)\citenamefont
  {Gemelke}, \citenamefont {Sarajlic}, \citenamefont {Bidel}, \citenamefont
  {Hong},\ and\ \citenamefont {Chu}}]{Gemelke2005}%
  \BibitemOpen
  \bibfield  {author} {\bibinfo {author} {\bibfnamefont {N.}~\bibnamefont
  {Gemelke}}, \bibinfo {author} {\bibfnamefont {E.}~\bibnamefont {Sarajlic}},
  \bibinfo {author} {\bibfnamefont {Y.}~\bibnamefont {Bidel}}, \bibinfo
  {author} {\bibfnamefont {S.}~\bibnamefont {Hong}}, \ and\ \bibinfo {author}
  {\bibfnamefont {S.}~\bibnamefont {Chu}},\ }\href {\doibase
  10.1103/PhysRevLett.95.170404} {\bibfield  {journal} {\bibinfo  {journal}
  {Phys. Rev. Lett.}\ }\textbf {\bibinfo {volume} {95}},\ \bibinfo {pages}
  {170404} (\bibinfo {year} {2005})}\BibitemShut {NoStop}%
\bibitem [{Note1()}]{Note1}%
  \BibitemOpen
  \bibinfo {note} {This is only approximately true, since the continued
  propagation in the harmonic confinement imparts different momentum
  corrections to each cloud.}\BibitemShut {Stop}%
\bibitem [{\citenamefont {Modugno}\ \emph {et~al.}(2003)\citenamefont
  {Modugno}, \citenamefont {Modugno}, \citenamefont {Roati}, \citenamefont
  {Fort},\ and\ \citenamefont {Inguscio}}]{Modugno2003}%
  \BibitemOpen
  \bibfield  {author} {\bibinfo {author} {\bibfnamefont {M.}~\bibnamefont
  {Modugno}}, \bibinfo {author} {\bibfnamefont {G.}~\bibnamefont {Modugno}},
  \bibinfo {author} {\bibfnamefont {G.}~\bibnamefont {Roati}}, \bibinfo
  {author} {\bibfnamefont {C.}~\bibnamefont {Fort}}, \ and\ \bibinfo {author}
  {\bibfnamefont {M.}~\bibnamefont {Inguscio}},\ }\href {\doibase
  10.1103/PhysRevA.67.023608} {\bibfield  {journal} {\bibinfo  {journal} {Phys.
  Rev. A}\ }\textbf {\bibinfo {volume} {67}},\ \bibinfo {pages} {023608}
  (\bibinfo {year} {2003})}\BibitemShut {NoStop}%
\end{thebibliography}%

\end{document}